\documentclass[prb,amsmath,twocolumn,final,superscriptaddress,10pt,aps]{revtex4-1}
\usepackage{graphicx}
\usepackage[utf8]{inputenc}

\newcommand{\be}{\begin{equation}}
\newcommand{\ee}{\end{equation}}
\newcommand{\quant}[2]{$#1\,\text{#2}$}

\newcommand{\bibspath}{/home/mleitner/aktiv/text/bibtex-referenzen/}
\newcommand{\eva}{\epsilon^\text{v}_\alpha}
\newcommand{\evb}{\epsilon^\text{v}_\beta}
\newcommand{\eab}{\epsilon^\text{A}_\beta}
\newcommand{\eba}{\epsilon^\text{B}_\alpha}
\newcommand{\eanb}{\epsilon^A_{\alpha\rightarrow\beta}}
\newcommand{\ebna}{\epsilon^B_{\beta\rightarrow\alpha}}

\begin{document}

\title{Thermodynamics of point defects and diffusion mechanisms in B2-ordered compounds}

\author{Michael Leitner}
\email{michael.leitner@frm2.tum.de}
\affiliation{Universität Wien, Fakultät für Physik, Boltzmanngasse 5, 1090 Wien, Austria}
\affiliation{Heinz Maier-Leibnitz Zentrum (MLZ), Technische Universität München, Lichtenbergstr.\ 1, 85748 Garching, Germany}
\date{\today}


\begin{abstract}
The point defect thermodynamics in a general family of binary compounds, including B2 compounds as a specific representative, are classified by way of two non-trivial energy parameters. The scheme is applied to published ab initio defect formation energies, and the variety of resulting phenomena is demonstrated. Further, by introducing model assumptions the consequences for the active diffusion mechanisms are deduced. It is shown that particularly for the off-stoichiometric case, the assumed prevalence of either the six-jump cycle or the triple-defect mechanism has to be reconsidered, as a number of qualitatively different mechanisms emerge as likely candidates for the dominant effect. Two of those, the 4+2-jump cycles and the waltzing-step mechanism, are introduced here.
\end{abstract}


\maketitle
\section{Introduction}
In pure elemental crystalline systems, the question of the thermal point defects' nature as well as of the active mechanisms for self diffusion can be considered as largely solved. Specifically for metals, the native point defect is the vacancy, whose random, thermally activated walk is also responsible for the stochastic translation of the atoms leading to self diffusion, while for semiconductors also diffusion via self-interstitials is plausible. For ordered compounds, the situation is much more complicated. Neglecting interstitials, already for a binary compound composed of two sublattices four fundamental point defects are possible, namely vacancies and antisite atoms on either sublattice. Further, the concentrations of these defects for a given system as a function of temperature are not independent variables, as experimentally the composition of the system has to be considered as fixed. Several theories to solve this issue have been put forward.\cite{hagenphilmaga1998,schottphysstatsolb1997,faehnleintermetallics1999}

The mechanisms responsible for diffusion in an ordered compound are still controversial. An uncorrelated random walk of a vacancy composed of nearest-neighbour jumps, as familiar from elemental metals, can be ruled out, as it would destroy the ordered structure during its migration. Specifically the B2-structure and analogous compounds, where a given atom's nearest neighbours are exclusively on the sublattice occupied by the other element, a vacancy could only move by jumps to farther neighbours within one sublattice without introducing antisite atoms. While such next-nearest neighbour jumps are a priori entirely possible, it was recognized early on that they would correspond to large migration energies. H.\ B.\ Huntington is attributed\cite{elcockpr1958} the first suggestion of a mechanism where a vacancy performs a correlated walk, so that it annihilates at later steps the antisite atoms created during earlier steps, leading to diffusion via nearest-neighbour jumps while conserving the atomic order in the long run. Apart from next-nearest neighbour jumps, these so-called six-jump cycles and the later proposal of the triple-defect mechanism\cite{stolwijkphilmaga1980} are the candidates that are conventionally considered for explaining experimental diffusion data\cite{mehrermattransjim1996} or for predicting the active mechanism from theoretical calculations\cite{xuprb2010}. However, it is conceivable that other mechanisms have hitherto been overlooked, which is sometimes conceded explicitly (e.g., Ref.\ \onlinecite{athenesphilmaga1997}).

The aim of this article is to explore the range of qualitatively different behaviours with respect to constitutional and thermal defect structures and, related to this issue, the distinct conceivable diffusion mechanisms, with special consideration to the B2 structure as one of the simplest intermetallic structures. It will recapitulate the results of a grand-canonical theory of defect concentrations, propose a re-parametrization of the effective defect formation energies as reported by theoretical calculations, derive the phase diagram of dominant constitutional and thermal defects with respect to these energy parameters, and classify published ab-initio defect energies for a variety of B2 systems accordingly. Further, an enumeration of the plausible diffusion mechanisms mediated by nearest-neighbour vacancy migration will be given, along with an evaluation of their respective energies for above parametrization under two simple models. This analysis will yield the surprising result that, in contrast to prevailing assurance, neither six-jump cycles nor the triple-defect mechanism are expected to be responsible for diffusion in many prominent systems, but one of two qualitatively different mechanisms proposed here. Finally, extant knowledge on diffusion in NiAl and CoGa will be discussed in the light of the insights outlined above.

\section{Point defect thermodynamics}
\subsection{Grand-canonical theory}\label{grcantheory}
Here the main results of Fähnle and coworkers' grand-canonical theory for the point defect concentrations in ordered structures will be recapitulated (for a thorough account see, e.g., Refs.\ \onlinecite{faehnleintermetallics1999} and \onlinecite{meyerprb1999} and the references cited therein). It deals with the point defect concentrations in ordered structures neglecting defect interactions, in its initial formulation for an ordered binary structure of AB stoichiometry with two sublattices. Strictly speaking, the latter condition can be relaxed to require only chemical, but not crystallographic equivalence between one constituent's native sites. This encompasses most of \textit{Strukturbericht} type B structures such as B1 (NaCl), B2 (CsCl), B3 (ZnS), or L1$_0$ (CuAu), to name only the most prominent, along with B20 (FeSi) as an example with eight sublattices, which, however, correspond to only two chemically inequivalent sites. While quantitative generalizations for situations with higher complexity, such for as D$0_3$, have been given \cite{besterprb1998}, the qualitative effects can directly be derived from the special case treated here. Note also that while there exist generalizations to treat interactions between defects in an approximate way \cite{semenovasolidstatesci2008}, neglecting interactions is exact in the limit of small defect concentrations as long as no complex defects appear and the system is thermodynamically stable (i.e., no phase separation by defect agglomeration occurs).

Four basic point defects are considered, to wit vacancies or anti-structure atoms on either sublattice (interstitials are neglected for simplicity, as their equilibrium concentration has been recognized to be negligible in comparison to vacancies and antisites in intermetallic compounds). The salient point of the theory is that in ordered compounds the notion of a given basic point defect's formation energy is ill-defined, as due to particle number conservation any such defect has to be generated in conjunction with at least another one. Therefore, one chooses a grand-canonical setting and considers so-called grand-canonical defects, where the particle numbers are not conserved and whose formation energies are defined up to one degree of freedom that has the role of a chemical potential. For any given temperature this chemical potential is implicitly defined so as to give the correct composition. 

The main result is that in the non-degenerate low-temperature limit a single type of defect dominates in the off-stoichiometric case (termed the constitutional defect). Therefore the chemical potential converges to some specific value, and the equations describing the concentration of the various basic defects simplify to Arrhenius-type expressions defined by so-called effective formation energies, which do not depend on temperature or composition. However, the pre-factor of these expressions includes a concentration-dependent term in addition to the conventional electronic and vibrational contributions to defect formation entropies. Analogously, in the case of perfect stoichiometry two basic point defects appear in coupled concentrations as the dominant low-temperature thermal defects, and the same simplifications ensue. For compositions and temperatures where the numbers of constitutional and thermal point defects are comparable, the full set of coupled non-linear equations has to be solved. In the following it is assumed that either constitutional or thermal defect concentrations dominate over the other. 

To define the nomenclature, let A and B be the constituent elements, with their native sublattices $\alpha$ and $\beta$. Naturally, an $\alpha$ vacancy is an empty site on the $\alpha$ sublattice, and an $\alpha$ antisite is a site on the $\alpha$ sublattice occupied by a B atom, which will be called B anti-structure atom. To motivate the behaviour reported above, define $\eva+\mu$ as the grand-canonical energy difference when removing one A atom from its native $\alpha$-sublattice, analogously $\evb-\mu$ for removing B from $\beta$, further $\eba+2\mu$ and $\eab-2\mu$ as the energy differences for replacing one A atom on its native sublattice by B and vice versa, respectively. Note that when calculating a given configuration's total energy, the term proportional to $\mu$ indeed depends only on the composition, so that introducing any number of defects while keeping the particle number constant is independent of $\mu$, the chemical potential for exchanging B for A. Now choose $\mu$ so that each of above energy differences is positive (this is possible, as otherwise the ordered structure would be thermodynamically unstable). If now temperature is lowered for fixed $\mu$, the composition of the system evolves towards stoichiometry (each defect gets progressively less probable). To keep the system at a certain off-stoichiometric composition, $\mu$ has to be adjusted in step with temperature (clearly the composition varies monotonically with $\mu$, so there is a unique solution). It is now obvious that for a certain off-stoichiometric composition there has to be a defect whose energy goes to zero as temperature goes to zero. This defect is the constitutional defect with effective formation energy equal to zero, and the other defects' effective formation energies follow from the corresponding $\mu$. In the case of stoichiometry on the other hand, for a general choice of $\mu$ one defect will have the smallest formation energy, so its concentration will dominate over the other defects for small temperature, which violates the condition of stoichiometry. As a consequence, in this case $\mu$ has to be chosen so that the minimal effective formation energy is shared by two complementary defects (i.e., either both vacancies or both antisites, or $\alpha$ vacancies and $\beta$ antisites or vice versa).

\subsection{Defect formation energy parametrization and phase diagrams of defect types}\label{parametrization}
\begin{figure*}
\includegraphics{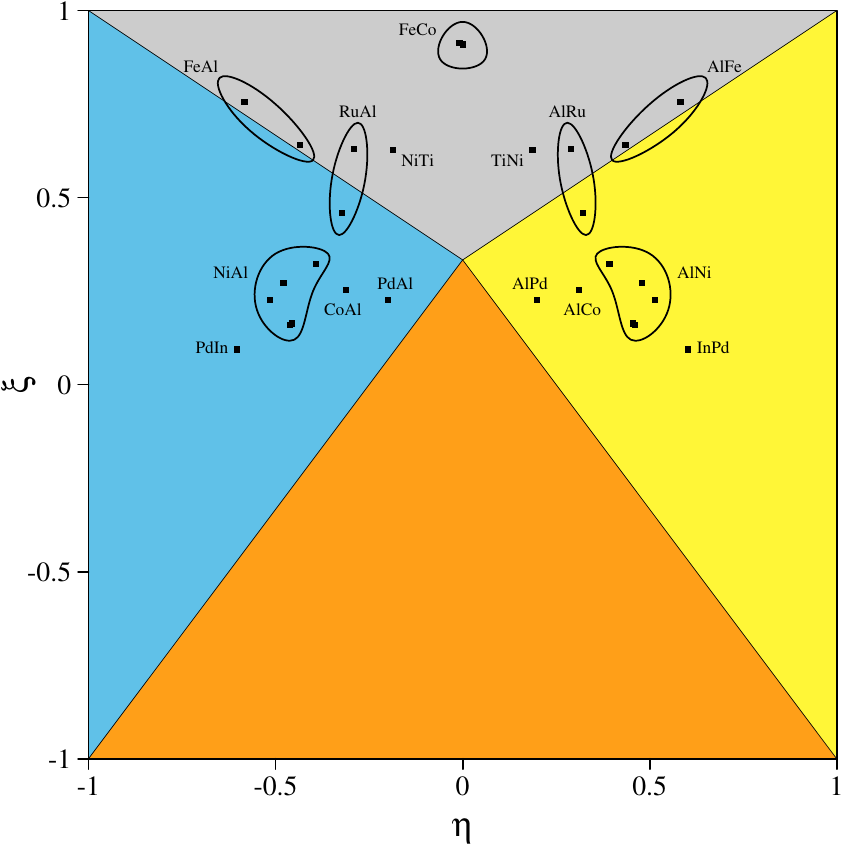}\qquad\includegraphics{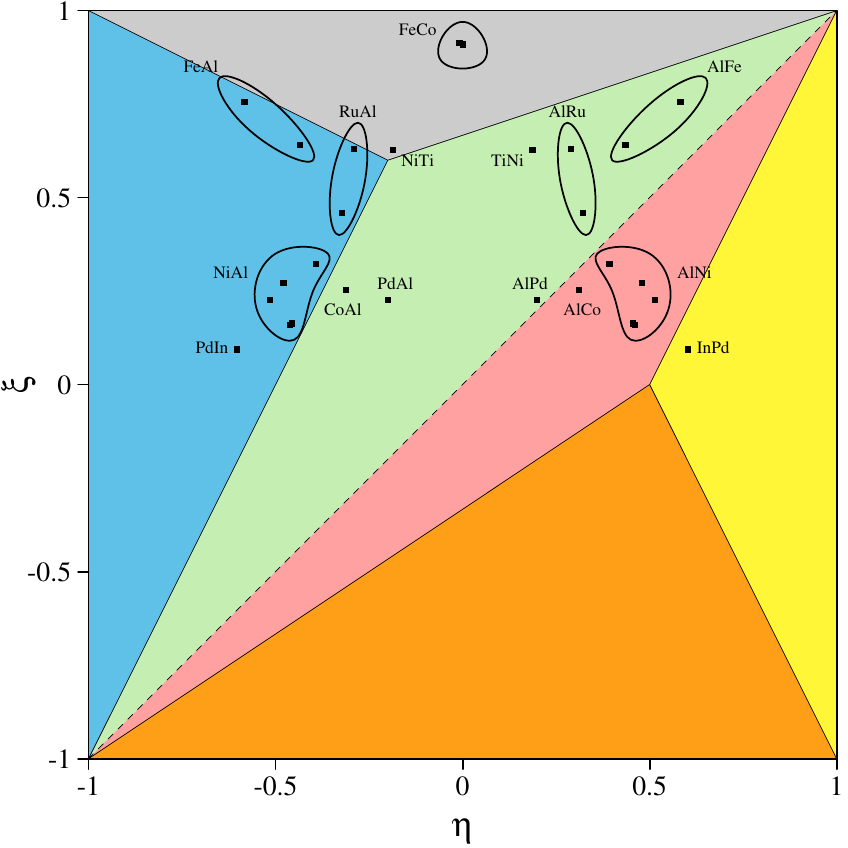}
\caption{Phase diagrams of constitutional and dominant thermal point defects of an AB-system for stoichiometry (left) and A-excess (right), where the dashed line separates the regions of constituional antisites (above diagonal) and vacancies (below). Dominating thermal excitations are coded by colour, where orange and gray are vacancy and antisite pairs, blue and yellow are A and B triple defects, and green and pink are antisite and vacancy annihilation, respectively.}\label{phdg_def}
\end{figure*}
This section proposes a reparametrization of the defect formation energies in order to explore the range of resulting defect structures. To this end define
\begin{subequations}\label{definitionparameter}
\begin{align}
E_0&=\eva+\evb+(\eab+\eba)/2\\
\xi&=\bigl((\eva+\evb)-(\eab+\eba)/2\bigr)/E_0\\
\eta&=\bigl((\eva-\evb)+(\eab-\eba)/2\bigr)/E_0.
\end{align}
\end{subequations}
Note that for computing these quantities, the grand-canonical energy differences for any choice of chemical potential $\mu$ can be used, as the corresponding contribution cancels in each case. Therefore, these three parameters define the defect energetics of a given system in a unique and non-redundant way. The positive parameter $E_0$ has the dimension of an energy and its only action corresponds to a rescaling of temperature, while $\xi$ and $\eta$ are dimensionless and vary from $-1$ to $1$. In the following, the constitutional and thermal defect structures correspond to a given choice of $(\xi,\eta)$ will be explored.

For the stoichiometric case, the four possible thermal excitations of the site occupations are (1) a vacancy pair, wherein the crystal is enlarged by one cell, generating one $\alpha$ and one $\beta$ vacancy, (2) an antisite pair, wherein an A atom is swapped with a B atom, generating one $\alpha$ and one $\beta$ antisite, (3) a triple defect in A, wherein the crystal is enlarged by one cell and an A atom is moved into the resulting $\beta$ vacancy, generating two $\alpha$ vacancies and one $\beta$ antisite, and lastly (4) analogously a triple defect in B. In the case of vanishing defect concentrations, the configurational entropy corresponding to one of above excitations is equal to the number of affected sites times $k_\text{B}\log(N)$, where $N$ is the number of sites per sublattice (this constant factor will be suppressed in the following). The formation energies and entropies are tabulated in Tab.\ \ref{stoideftab}. The dominant thermal excitation resulting for a specific $(\xi,\eta)$ is the one with the smallest cost in energy per gain in entropy, as given in the last column of the table. The corresponding phase diagram is drawn in Fig.\ \ref{phdg_def} (left), with the single critical point being at $(1/3,0)$. 

\begin{table}
\begin{tabular}{c|rcc}
excitation&energy $E$&entropy $S$&$E/SE_0$\\
\hline
vacancy pair&$\eva+\evb$&$2$&$(1+\xi)/4$\\
antisite pair&$\eab+\eba$&$2$&$(1-\xi)/2$\\
A triple defect&$2\eva+\eab$&$3$&$(1+\eta)/3$\\
B triple defect&$2\evb+\eba$&$3$&$(1-\eta)/3$
\end{tabular}
\caption{Formation energies, entropies and formation energy per entropy increase of thermal occupational excitations for stoichiometry}\label{stoideftab}
\end{table}

For off-stoichiometry only the case of A-excess shall be treated, as the results for B-excess follow by symmetry. Here the primary distinction is the nature of the constitutional defects. An excess of two A atoms can be accommodated either by two $\beta$ vacancies or one $\beta$ antisite. Therefore, if $2\evb-\eab=E_0(\xi-\eta)>0$ the constitutional defect will be the $\beta$ antisite, otherwise the $\beta$ vacancy. The question of the thermal excitations is analogous to the stoichiometric case with the following complications: First, in addition to above four excitations there exists in the case of constitutional antisites the possibility of (5) antisite annihilation, wherein the crystal is enlarged by one cell and one constitutional A anti-structure atom is moved into the $\alpha$ vacancy, generating two $\beta$ vacancies and annihilating one $\beta$ antisite, and in the case of constitutional vacancies analogously the possibility of (6) vacancy annihilation, wherein one A atom moves into a constitutional $\beta$ vacancy and the crystal is shrunk by one cell. These additional excitations have been called interbranch defects by Korzhavyi et al. \cite{korzhavyiprb2000}. Second, as the thermal defect concentration is by assumption much smaller than the constitutional defect concentration, the thermal excitation (or annihilation) of point defects that exist already as constitutional defects does not affect the entropy of the system. The relevant expressions are given in Tab.\ \ref{offstoideftab}.
\begin{table}
\begin{tabular}{c|rcc}
excitation&energy $E$&entropy $S$&$E/SE_0$\\
\hline
vacancy pair&$\eva+\evb$&$2$&$(1+\xi)/4$\\
antisite pair&$\eab+\eba$&$1$&$1-\xi$\\
A triple defect&$2\eva+\eab$&$2$&$(1+\eta)/2$\\
B triple defect&$2\evb+\eba$&$3$&$(1-\eta)/3$\\
antisite annihil.&$2\evb-\eab$&$2$&$(\xi-\eta)/2$
\end{tabular}\vspace{.5cm}
\begin{tabular}{c|rcc}
excitation&energy $E$&entropy $S$&$E/SE_0$\\
\hline
vacancy pair&$\eva+\evb$&$1$&$(1+\xi)/2$\\
antisite pair&$\eab+\eba$&$2$&$(1-\xi)/2$\\
A triple defect&$2\eva+\eab$&$3$&$(1+\eta)/3$\\
B triple defect&$2\evb+\eba$&$1$&$1-\eta$\\
vacancy annihil.&$-2\evb+\eab$&$1$&$-\xi+\eta$
\end{tabular}
\caption{Formation energies, entropies and formation energy per entropy increase of thermal occupational excitations for A-excess, for the cases of constitutional $\alpha$ antisites (above) or $\beta$ vacancies (below).}\label{offstoideftab}
\end{table}
Note that from the argumentation in Sect.\ \ref{grcantheory} it follows that the dominant thermal point defect is the one with the lowest effective formation energy  after the constitutional point defect with vanishing effective formation energy. Therefore only those thermal excitations that introduce just one kind of thermal point defect in addition to the constitutional defect can be expected to dominate. This is in agreement with the dominant excitations that follow from Tab.\ \ref{offstoideftab}: For instance, in the case of constitutional antisites $E/S$ for the vacancy pair is equal to the mean values of $E/S$ for A triple defects and antisite annihilation, so one of those has smaller $E/S$ and vacancy pairs will hence not occur for any choice of $(\xi,\eta)$. The resulting phase diagram is drawn in Fig.\ \ref{phdg_def} (right). Here the critical points are given by $(3/5,-1/5)$ and $(0,1/2)$.

To demonstrate that the theoretical richness of defect thermodynamics is actually realized, published values of defect energetics in a range of B2 systems obtained by ab-initio density functional theory calculations have been compiled and are reported in Tab.\ \ref{compilation}. The qualities of the respective calculations differ, which is evidenced for instance by the scatter of the values corresponding to NiAl, a popular system for defect thermodynamics investigations. However, plotting the energy parameters in the defect structure phase diagrams of Fig.\ \ref{phdg_def} shows that the differences between different systems are larger than these deviations, so that the qualitative conclusions seem sound. Note that the resulting classification is as exact as the input from the ab-initio calculations.

\begin{table}
\begin{tabular}{r|ccccccc|c}
&$\eva$&$\evb$&$\eab$&$\eba$&$E_0$&$\xi$&$\eta$&Ref.\\
&(eV)&(eV)&(eV)&(eV)&(eV)&&&\\
\hline
FeAl&$0.81$&$4.01$&$0.67$&$0.67$&$5.49$&$0.756$&$-0.583$&\onlinecite{fuprb1995}\\
&$1.06$&$3.46$&$0.99$&$0.99$&$5.51$&$0.641$&$-0.436$&\onlinecite{besterprb1999}\\
\hline
RuAl&$1.46$&$3.07$&$1.03$&$1.03$&$5.56$&$0.629$&$-0.290$&\onlinecite{sotmatscipol2005}\\
&$1.45$&$3.21$&$1.45$&$2.02$&$6.39$&$0.458$&$-0.321$&\onlinecite{prinsactamat2007}\\
\hline
CoAl&$1.29$&$2.23$&$1.29$&$2.91$&$5.62$&$0.253$&$-0.311$&\onlinecite{besterprb1999}\\
\hline
NiAl&$0.94$&$2.13$&$0.94$&$2.21$&$4.64$&$0.322$&$-0.393$&\onlinecite{fuprb1995}\\
&$0.74$&$1.97$&$0.74$&$2.36$&$4.26$&$0.272$&$-0.479$&\onlinecite{besterprb1999}\\
&$0.79$&$1.74$&$0.79$&$2.85$&$4.35$&$0.163$&$-0.455$&\onlinecite{korzhavyiprb2000}\\
&$0.55$&$1.53$&$0.55$&$2.08$&$3.40$&$0.225$&$-0.514$&\onlinecite{jiangactamat2005a}\\
&$0.73$&$1.62$&$0.73$&$2.68$&$4.05$&$0.159$&$-0.461$&\onlinecite{prinsactamat2007}\\
\hline
PdAl&$1.22$&$1.58$&$1.22$&$2.31$&$4.56$&$0.227$&$-0.198$&\onlinecite{fuprb1995}\\
\hline
PdIn&$0.31$&$0.97$&$0.31$&$1.81$&$2.34$&$0.094$&$-0.603$&\onlinecite{jiangintermet2006}\\
\hline
NiTi&$1.09$&$1.74$&$0.65$&$0.65$&$3.48$&$0.626$&$-0.187$&\onlinecite{luprb2007}\\
\hline
FeCo&$1.65$&$1.68$&$0.15$&$0.15$&$3.48$&$0.914$&$-0.009$&\onlinecite{neumayerprb2001}\\
&$1.78$&$1.78$&$0.17$&$0.17$&$3.73$&$0.909$&$0.000$&\onlinecite{fuprb2006}\\
\end{tabular}
\caption{Effective defect formation energies for stoichiometry in AB nomenclature as taken or computed from published ab-initio values, along with corresponding energy parameters.}\label{compilation}
\end{table}

The two panels of Fig.\ \ref{phdg_def} show that indeed all principally possible defect structures are realized for some $(\xi,\eta)$: For the stoichiometric case the thermal excitation has to be composed of one point defect that in itself would lead to A-excess and one to B-excess. Of each there are two possibilities, so there result four distinct thermal excitations. On the other hand, the case of off-stoichiometry corresponds to one point defect having zero effective formation energy, so treating without loss of generality only A-excess, two choices are possible. For each of these, one of the remaining three point defects will have the lowest effective formation energy and therefore dominate, giving six possible cases of defect structures. What is remarkable is that, with the exception of vacancy pairs, all these options are realized already by the small subset of intermetallic systems in Tab.\ \ref{compilation}. For compounds of increasingly ionic character, vacancies will decrease in formation energy compared to antisite defects, which would favour thermal vacancy pairs. In fact, for ionic compounds with comparative sizes of cations and anions, such as for the prototypical CsCl, the Schottky defect (i.e., vacancy pairs) is known to be the dominant thermal excitation\cite{harveyphilmag1967}.

It is customary to classify the thermodynamics of AB compounds into antistructure-type systems and triple-defect systems. This is due to Neumann\cite{neumannactamet1980}, who observed the dominant defect type to be connected to the formation enthalpy of the compound. Hybrid behaviour was predicted for borderline cases such as FeAl. In further work, the distinction was interpreted as the opposite extremes of parameter choices in general thermodynamical models (e.g, Ref.\ \onlinecite{krachlerjphyschemsolids1989}), with the label ``hybrid behaviour'' assigned to intermediate cases. In most cases, it is understood that the distinction refers to the nature of constitutional defects, where ``anti-structure system'' means that deviations in both directions from stoichiometry are accomodated by constitutional antisites, while in ``triple-defect systems'' only, say, A-excess is accomodated by A anti-structure atoms, while A-deficiency is accomodated by $\alpha$ vacancies. However, sometimes it is also implied that, e.g., in anti-structure systems not only the constitutional defects, but also the ``governing'' thermal defects are antisites (equally for triple-defect systems)\cite{breuerphilmaga2002,stolwijkphilmaga1980}, while in other cases the distinction is assumed to be defined directly in terms of the dominant thermal excitations instead of the way off-stoichiometry is accomodated \cite{bestermatscienga2002}. 

Fig.\ \ref{phdg_def} shows that care has to be taken, as above definitions are by no means equivalent or sometimes even well-defined: First, the dominant thermal excitation depends on whether the composition is A-deficient, stoichiometric, or A-excessive. Therefore, there are two different consistent possibilities to interpret a given system being classified as a triple-defect system in terms of the dominant thermal excitation, namely either only for the stoichiometric case, corresponding to $|\eta|>\max(6\xi-2, -3\xi+1)/4$, or for all cases, corresponding to $|\eta|>(1+|\xi|)/2$. In contrast, the common understanding of constitutional antisites on one side of stoichiometry and constitutional vacancies on the other would correspond to $|\eta|>|\xi|$. Such a distinction in terms of constitutional defects seems most natural, so that calling this situation the triple-defect case is unfortunate, as a triple defect is always a thermal excitation. It is conceivable that this issue is partly responsible for the inconsistent interpretations. Also, claims of hybrid behaviour, e.g., for FeAl can have resulted from triple defects being the dominant thermal excitation for Fe-rich compositions, while off-stoichiometry is accomodated by antisites, so that the most abundant thermal point defect is the Fe vacancy. Concludingly, it seems most expedient to use a top-level classification of AB compounds according to the constitutional defects of anti-structure accomodation with $\xi>|\eta|$, mixed accomodation (anti-structure atoms on one side of the stoichiometry, vacancies on the other) for $|\eta|>|\xi|$ and vacancy accomodation for the hypothetical case of $\xi<-|\eta|$, which apparently has not been reported yet.

The effective defect formation energies are given by
\begin{subequations}\label{effformenerg}
\begin{align}
\eva/E_0&=1/4+\xi/4+2\eta/5+\mu\\
\evb/E_0&=1/4+\xi/4-2\eta/5-\mu\\
\eba/E_0&=1/2-\xi/2-\eta/5+2\mu\\
\eab/E_0&=1/2-\xi/2+\eta/5-2\mu
\end{align}
\end{subequations}
with
\begin{subequations}\label{mufaelle}
\begin{align}
\text{vacancy pairs:}&&\mu&=-2\eta/5\\
\text{antisite pairs:}&&\mu&=\eta/10\\
\text{A triple defects:}&&\mu&=(1/4-3\xi/4-\eta/5)/3\\
\text{B triple defects:}&&\mu&=(-1/4+3\xi/4-\eta/5)/3
\end{align}
for the regions of different dominant thermal excitations in stoichiometry and
\begin{align}
\text{structural antisites:}&&\mu&=1/4-\xi/4+\eta/10\\
\text{structural vacancies:}&&\mu&=1/4+\xi/4-2\eta/5
\end{align}
\end{subequations}
for A-rich off-stoichiometry. Note that always the mirror case for a given expression (i.e., exchanging the roles of A and B as well as those of $\alpha$ and $\beta$) can be obtained by flipping the signs of $\mu$ and $\eta$. It is easy to verify that Eqs.~\eqref{definitionparameter} are consistent with Eqs.~\eqref{effformenerg}, and that Eqs.~\eqref{mufaelle} indeed lead to vanishing effective formation energies of the respective constitutional defects in the off-stoichiometric case and to equal effective formation energies of the primitive defects making up the dominant thermal excitations in the stoichiometric case. 

It is worth noting that for a given system not necessarily all primitive defects actually exist. For instance, in NiAl the Al vacancy is separated only by a barrier of \quant{0.35}{eV} against the decay into a Ni vacancy and a Ni anti-structure atom (see Sect.\ \ref{nial}). For higher asymmetries, the barrier could vanish, so that the isolated vacancy would exist only as an extremely anharmonic oscillation on phonon timescales. However, for the present purposes it would be correct to use the energy of the nearest-neighbour vacancy-antisite pair in place of the unstable vacancy, with an appropriate interpretation of the resulting point defect concentrations.

Finally, for most systems of interest the criteria for the stoichiometric case, i.e., that thermal defects are much more abundant than constitutional defects, are quite stringent. Take for instance NiAl with an effective formation energy of thermal antisites around \quant{0.75}{eV} according to Tab.\ \ref{compilation}. For an intermetallic compound deviations of at least $0.1 \%$ from nominal stoichiometry have to be realistically expected in the experiment. Only at the congruent melting temperature of \quant{1911}{K} the thermal defect concentration reaches 1\%, so that the stoichiometric limit can be considered as valid. In contrast, the off-stoichiometric limit will be appropriate for the majority of experimental cases in well-ordered systems. Only for weakly-ordered systems such as FeCo, which display an order-disorder transition within the solid phase, large concentrations of thermal defects will exist at intermediate temperatures.

\section{Diffusion}
\subsection{Transition state theory}\label{transstatetheory}
Experiment has proven that the hopping rate, and as a consequence the diffusion constant, of a single jump diffusion mechanism such as vacancy diffusion in elemental crystals can often be described well by an Arrhenius-type expression. Indeed, transition state theory as formulated by Vineyard \cite{vineyardjphyschemsolids1957} leads to exactly such an expression for the hopping rate, where the activation energy is equal to the difference between saddle point energy and initial energy in the $3N$-dimensional energy landscape of the $N$-atom subsystem considered explicitly, and the frequency prefactor is the product of all $3N$ eigenfrequencies in the initial stable state divided by the $3N-1$ real eigenfrequencies in the saddle point, provided the harmonic approximation is valid in both states. In addition to this condition, whose failure can result in deviations from Arrhenius behaviour, transition state theory assumes that once a system crosses over the saddle point, it continues until the neighbouring local minimum, where it thermalizes. If the time scales of the jump duration and the mean residence time in a local minimum differ enough and the saddle points in $3N$-dimensional space are narrow enough so that an immediate further jump is improbable, then $N$ can be chosen so that the subsystem behaves microcanonically during the jump, i.e., does not reverse, while the coupling to the heat bath constituted by the other atoms leads to thermalization in the final local minimum. These conditions are normally fulfilled for bulk vacancy diffusion.

As discussed below, in compounds often correlated jump sequences are relevant for diffusion. There, the system has to pass through a number of local energy minima, called stable states from now on, for completing the jump sequence. In each of these it will rest long enough for thermalizing, and therefore can potentially pass back over the energy apex along the jump path. As a consequence, the above account has to be extended for treating diffusion in compounds.

Here the comparatively simple case of diffusion via a unique string of stable states will be considered. Let the initial stable state be the energy minimum along the string, called the ground state, and denote its energy by $E^\text{m}_0$. It is separated by a saddle point with energy $E^\text{s}_1$ from the next stable state with energy $E^\text{m}_1$, up to the final stable state $M$, with the corresponding products of eigenfrequencies $\omega^\text{m}_i$ and $\omega^\text{s}_i$. According to transition state theory the flux between states $i$ and $i+1$ is given by
\be
\Phi_{i\rightarrow i+1}=\frac{c_i\omega^\text{m}_ie^{-E^\text{m}_i/k_\text{B}T}-c_{i+1}\omega^\text{m}_{i+1}e^{-E^\text{m}_{i+1}/k_\text{B}T}}{\omega^\text{s}_{i+1}e^{E^\text{s}_{i+1}/k_\text{B}T}},\label{flusseins}
\ee
where $c_i$ is the concentration in the respective stable states. For obtaining the steady-state flux identify all fluxes with $\Phi$, solve Eq.\ \eqref{flusseins} for $c_{i+1}$ and perform repeated substitution, yielding
\be
\Phi=\frac{c_0\omega^\text{m}_0e^{-E^\text{m}_0/k_\text{B}T}-c_M\omega^\text{m}_Me^{-E^\text{m}_M/k_\text{B}T}}{\sum\omega^\text{s}_i e^{E^\text{s}_i/k_\text{B}T}}.\label{flussviele}
\ee
Comparing Eqs.\ \eqref{flusseins} and \eqref{flussviele} shows that the simple Arrhenius expression for diffusion via direct jumps between stable states has to be replaced by the reciprocal sum of such expressions for diffusion via intermediate stable states. However, for realistic choices of $\omega^\text{s}_i$ and $E^\text{s}_i$ the errors introduced by assuming a simple Arrhenius behaviour with the highest saddle point energy as activation energy are minute, specifically it is exact in the small-temperature limit. This qualitative conclusion agrees with detailed calculations of, e.g., the six-jump cycles in B2-compounds, where there are several paths connecting initial and final state \cite{aritaactamet1989,drautzactamat1999}. 




\subsection{Diffusion mechanisms}\label{mechanisms}
Arguably the simplest class of ordered compounds is constituted by the binary compounds of AB stoichiometry where all nearest-neighbour sites of an $\alpha$ site are $\beta$ sites and vice versa. Here a list of conceivable diffusion mechanisms for these compounds will be given. In Sect.\ \ref{completeness} it will be argued that under plausible assumptions this list is exhaustive. A two-dimensional square lattice with alternating occupations will be used as a model for illustrating the mechanism. For actual intermetallic compounds, the B2 structure is by far the most prominent representative of this class of compounds, and special consideration will be devoted to it below.

Historically, with the growing acceptance of stochastic nearest-neighbour vacancy jump diffusion being the dominant mechanism of diffusion in elemental metals, it was recognized that such a mechanism is incompatible with the observed stability of the ordered structure during diffusion \cite{lidiardpr1957}. Apart from the possibility of any element performing jumps directly within its native sublattice, corresponding to jumps to second-nearest neighbour sites, correlated jump mechanisms composed of nearest-neighbour jumps have been proposed to obviate the high migration energies assumed for these farther jumps. These correlated jump sequences are the six-jump cycles \cite{elcockpr1958} and the divacancy mechanism as believed to be active in ionic compounds, where it suggests itself already because of its favourable Coulomb energy. In the case of intermetallic compounds, the divacancy mechanism is often called triple-defect mechanism \cite{stolwijkphilmaga1980}, as for large thermodynamical asymmetries a localized triple defect as opposed to the divacancy can be the ground state for this jump mechanism. At large deviations from stoichiometry diffusion via percolating antisite defects can become possible, termed the anti-structure bridge mechanism \cite{kaointermet1993}. 

The predominant opinion today is that diffusion will happen via the anti-structure bridge mechanism if possible, otherwise via one of the correlated jump mechanisms, i.e., the six-jump cycle or the triple-defect mechanism\cite{baloghdiffusion2014}. While according to calculations the migration energies for jumps between second-nearest neighbours are not as high as to be ruled out a priori (compare, e.g., Ref.\ \onlinecite{xuintermet2009} for NiAl, which is arguably the most prominent system for these kinds of studies), the constituents' respective activation energies of diffusion often seem to be coupled, which hints towards correlated jump mechanisms. The following systematic discussion will be restricted to jumps between nearest-neighbour sites at low temperatures so that no thermal disorder has to be considered and differentiating between clear-cut diffusion mechanisms is possible.

\begin{figure}
\includegraphics{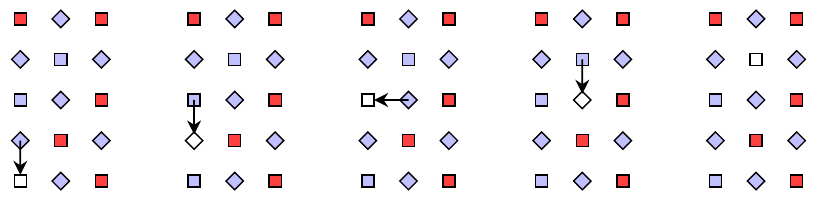}
\caption{Diffusion on percolating antisite cluster --- anti-structure bridge mechanism. $\alpha$ sites have diamond shape, $\beta$ sites have square shape; A atoms are blue, B atoms red, vacant sites are white. The thermal vacancy's movement via $\alpha$ sites and constitutional $\beta$ antisites leads to diffusion of the majority A component.}\label{asbridge}
\end{figure}

The first qualitative distinction is whether the mechanism can be active in an ideally ordered lattice or whether it needs percolating constitutional defects. For the latter possibility in the case of constitutional antisites this is the anti-structure bridge mechanism, where a vacancy jumps from the majority sublattice onto a constitutional antisite and back onto the majority sublattice. It is illustrated in Fig.\ \ref{asbridge} for A-excess (in the following called $\beta$ anti-structure bridge mechanism, due to the $\beta$ antisites acting as bridges). Evidently no additional disorder is introduced, therefore this mechanism is commonly assumed to have a small activation energy and be the dominant mechanism responsible for diffusion of the majority component once percolation permits. Ref.\ \onlinecite{divinskijphyscondmat1997} reports it to set in for the B2 structure at a composition of around 54.9(2)\% of A atoms (i.e., where 10\% of $\beta$ sites are occupied by A atoms). A more precise value for the percolation threshold in the academic case of ideal random occupancies corresponding to no defect interactions is 54.882225(5)\% \cite{*[{R. M. Ziff and S. Torquato as reported by }] [{}] kurzawskirepmathphys2012}.

\begin{figure}
\includegraphics{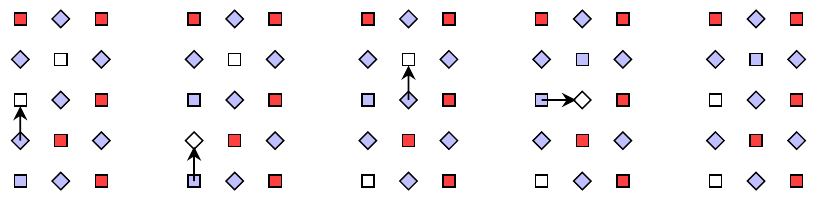}
\caption{Diffusion via percolating vacancy cluster --- vacancy bridge mechanism. Interpretation of symbols as in Fig.\ \ref{asbridge}. The movement of the thermal $\beta$ antisite between the constitutional $\beta$ vacancies leads to diffusion of the majority A component. The defect configuration and migration path mirror those of Fig.\ \ref{asbridge} to demonstrate the analogy.}\label{vacbridge}
\end{figure}

Also for the mirror situation of constitutional vacancies a low-energy percolation mechanism for majority species diffusion exists, which is illustrated in Fig.\ \ref{vacbridge}. For the case of constitutional $\beta$ vacancies, the diffusing defect is a thermal $\beta$ antisite. The elementary step here consists in a nearest neighbour of the anti-structure atom jumping into a constitutional vacancy, followed by the initial A anti-structure atom moving into the transient $\alpha$ vacancy. In effect, the thermal antisite defect migrates on the constitutional $\beta$ vacancies, just as in the case of the anti-structure bridge mechanism a thermal vacancy diffuses via constitutional antisites. The connectivity conditions are the same, so the same percolation threshold of 10\% vacancy concentration on the minority sublattice applies for the B2 structure. This corresponds to a deviation from stoichiometric composition of only 2.5\%, half of the value necessary for the anti-structure bridge mechanism. This mechanism has probably first been considered by Xu and van der Ven, who found it to be active in their simulations of Al-rich NiAl\cite{xuprb2010}. They called it anti-structure bridge sequence for Al atoms, which is unfortunate as it does not allow to distinguish this mechanism from a proper anti-structure bridge mechanism on thermal antisites as proposed by Herzig and Divinski for the same system \cite{herzigintermet2004}. Therefore, here it shall be termed vacancy bridge mechanism, as the constitutional vacancies provide the bridges for the diffusing thermal antisite defect, just as the constitutional antisites provide the bridges for the diffusing thermal vacancy in the anti-structure bridge mechanism. Note that the vacancy bridge mechanism can also be considered a variant of the classical interstitialcy mechanism of diffusion\cite{mccombiepr1956}, where the percolating vacant sites take the role of the interstitial sites' lattice.

For discussing diffusion mechanisms where a well-defined defect migrates in an otherwise ideal environment, it is useful to distinguish by way of the diffusing defect's charge, as this quantity is conserved over all stages of the mechanism. To this end assign the charge of $+1$ to an A atom irrespective of whether it is on its native sublattice or not, $0$ to a vacancy, and $-1$ to a B atom. In other words, the charge as defined here is the difference of the numbers of A and B atoms counted over the lattice cells that cover the migrating defect. 

For each defect charge, there exists a qualitatively unique string of states, i.e., assignments of atoms to lattice sites, linked by nearest-neighbour vacancy jumps, that generates the smallest number of antisite defects and that therefore is expected to correspond to the smallest migration energy. However, depending on the system, its actual realization can take a number of forms, distinguished by the geometry of the jump path, or which states along the path correspond to the ground and apex states (stable states lowest and highest in energy, respectively). Further, some of those states can actually be unstable, that is no local energy minimum, so that they are traversed by coupled jumps. The three simplest defects with the corresponding diffusion mechanisms are listed in the following, where for the two charged defects obviously also the mirrored case exists:

\begin{figure}
\includegraphics{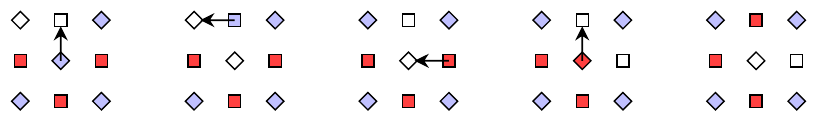}
\caption{Correlated propagation of neutral defect --- divacancy mechanism.}\label{divacancy}
\end{figure}

--- A neutral defect. Its movement is illustrated in Fig.\ \ref{divacancy}. It starts as a nearest-neighbour divacancy, then an A atom performs a nearest-neighbour jump into the $\beta$ vacancy, thereby creating a localized A triple defect. Successively, the A atom performs a further jump into the initial $\alpha$ vacancy, again leading to a divacancy. If now a B atom performs the analogous steps, the divacancy has effected a net movement. This is the classical divacancy mechanism\cite{dienesjchemphys1948}, but depending on the energetics, one of the triple defect states can be the ground state, for which case the term triple-defect mechanism has been coined\cite{stolwijkphilmaga1980}. For implying the opposite case with the divacancies being the ground state, leap-frog mechanism would be an appropriate name due to the movement of the two constituent vacancies. Independent of these distinctions, different jump geometries are possible, which in reality will differ in energy. Further, if (say) the A atom jumps are much more frequent than B atom jumps because the B triple defect is costly in energy, then the defect will oscillate rapidly between different orientations of the divacancy state and the A triple-defect state. During this time it will shuffle all neighbouring A atoms, before it concludes its movement by a B atom jump, thus leading to large diffusion asymmetries\cite{stolwijkphilmaga1980,bakkerphilmaga1981}.

\begin{figure}
\includegraphics{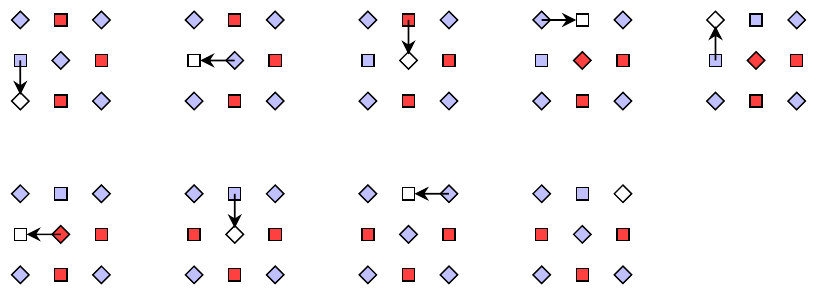}
\caption{Correlated propagation of single-charged defect --- jump cycle. If the ground states are the single $\beta$ vacancy states 2 and 8, the illustration depicts the classical six-jump cycle linking the ground states 2 and 8, framed by the last and first jump of the preceding and subsequent cycles, respectively. If on the other hand the states 1, 3, 7 and 9, each with an $\alpha$ vacancy and a $\beta$ antisite, are the ground states, it illustrates the 4+2-jump cycle with the introductory two-jump defect reorientation, the four-jump cycle, and a further subsequent defect reorientation.}\label{jumpcycle}
\end{figure}

--- A single-charged defect. Its movement by jump cycles is illustrated in Fig.\ \ref{jumpcycle} for positive charge, that is, starting with a $\beta$ vacancy, which in the following will be called $\beta$ jump cycles. Along a cycle of $K$ nearest-neighbour relations, the atoms successively move into the vacancy, until $K-1$ antisite defects are created. The vacancy continues on the cycle, so that those anti-structure atoms perform a second jump back onto their native sublattice. Finally, after $2K-2$ jumps only the initial vacancy remains, effectively translated by two nearest-neighbour jumps. Also $K/2$ B atoms and $K/2-1$ A atoms have been translated by the same amount. For the B2 structure, the smallest such cycle correspond to $K=4$, leading to the mechanism being called six-jump cycle and the initial estimate\cite{elcockpr1958} that the ratio of the two self-diffusion constants is at most 2. In contrast, for instance in the zincblende structure the smallest cycle consists of $K=6$ sites. For the square lattice as illustrated in Fig.\ \ref{jumpcycle} all such smallest cycles are equivalent, whereas for, e.g., the B2 structure there exist three geometrically inequivalent 4-site cycles. Also here different states can be the ground and apex states. Specifically, for significant asymmetries the second state with a $\beta$ antisite and an $\alpha$ vacancy can be lower in energy than the single $\beta$ vacancy. In this case the six-jump cycle would be more appropriately called 4+2-jump cycle, as the vacancy-antisite pair would be able to re-orient freely within its cell, until the vacancy eventually performs a cycle during which only two additional antisites are created. This had already been noted by Ath\`enes and coworkers\cite{athenesphilmaga1997}, but its relevance for actual systems as will be discussed in Sect.\ \ref{nial} has not been recognized yet.

\begin{figure}
\includegraphics{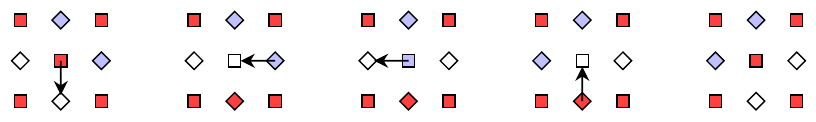}
\caption{Correlated propagation of doubly charged defect --- waltzing-step mechanism.}\label{walzer}
\end{figure} 

--- A double-charged defect. Its movement for the case of an initial $\alpha$ vacancy pair (called $\alpha$ waltzing-step mechanism) is illustrated in Fig.\ \ref{walzer}. It starts with a B atom moving into one of the $\alpha$ vacancies. Then an A atom performs a double jump via the vacated $\beta$ site into the other $\alpha$ vacancy, after which the B atom retakes its initial site. As a consequence, just as with the percolation mechanisms but in contrast to jump cycles and the divacancy mechanism, in this correlated mechanism only one constituent experiences a net translation. A fitting name for the described correlated movement is waltzing-step mechanism, as the B atom stepping temporarily out of the way for the A atom to pass perfectly mirrors the movement of the female and male partners during one bar of the classical ballroom dance. This mechanism has never been discussed yet, but below it will be shown that under reasonable assumptions it is a priori not less probable than for instance the divacancy mechanism.

\subsection{Completeness}\label{completeness}
Recapitulating the assumptions stated above, viz.\ a) small defect concentrations corresponding to a movement of the diffusing complex defect in a perfect lattice and b) this movement being composed of nearest-neighbour jumps into vacancies, above-discussed three qualitatively different mechanisms (divacancy mechanism, jump cycles, and waltzing-step mechanism) constitute an exhaustive list of possible candidates for the diffusion mechanism with lowest activation energy under the additional reasonable assumption that c) the introduction of additional transient disorder costs additional energy. This can be shown as follows: First, all three mechanisms constitute those sequences of jumps where the respective defects or defect complexes can migrate with the smallest amount of transient order. This claim can be formally verified by surveying the finite number of states that can be reached from the respective initial states under the constraint to not introduce more antisite disorder than in the candidate mechanisms. Further, allowing more than two vacancies does not allow the defect complex to migrate with less antisite disorder than both the divacancy and the waltzing-step mechanisms: The movement of a defect complex in an otherwise ideal lattice necessitates the movement of both constituents, at least a temporary stepping aside as in the waltzing-step mechanism. As an $\alpha$ site's direct neighbours are exclusively $\beta$ sites and vice versa, any movement of component A by nearest-neighbour jumps has to involve an $\alpha$ vacancy and a $\beta$ antisite, vice versa for B. The divacancy mechanism has only a single additional vacancy in the respective corresponding states. Naturally, its apex energy cannot be lowered by allowing additional vacancies. If the effective defect formation energies should show a large asymmetry, the waltzing-step mechanism can become active and obviate the more expensive transient triple defect, at the cost of an additional transient antisite defect. Introducing a third $\alpha$ vacancy (or even more) to an $\alpha$ waltzing-step mechanism does not obviate the transient $\alpha$ antisite, while the introduction of a $\beta$ vacancy allows the divacancy mechanism to proceed, where the second $\alpha$ vacancy can be eliminated. 


Of course, interactions between neighbouring defects as well as different energy barriers between the respective states can in principle enable mechanisms to be active that have not been considered above. Also, mechanisms involving interstitial sites (of which direct jumps to vacancies farther than on nearest-neighbour sites are the most simple representative) cannot be ruled out by such considerations. Finally, thermal disorder or significant constitutional defect concentrations below the percolation threshold will lead to a blurring of above clear-cut distinctions, where diffusion over small scales is promoted by analoga of the anti-structure or vacancy bridge mechanisms on non-percolating defect clusters, but a jump cycle or a cooperation of two vacancies is necessary for the diffusing defect to jump to another cluster.

\begin{table*}
\begin{tabular}{r|ccc}
&$\eta<-1/2+\xi/2$&$-1/2+\xi/2<\eta<1/2-\xi/2$&$1/2-\xi/2<\eta$\\
\hline
divacancy&$1+|\eta|$&$1+|\eta|$&$1+|\eta|$\\
$\alpha$ jump cycles&$7/4-5\xi/4-3\eta/5+\mu$&$7/4-5\xi/4-3\eta/5+\mu$&$5/4-3\xi/4+2\eta/5+\mu$\\
$\alpha$ waltzing-step&$1-\eta/5+2\mu$&$3/2-\xi/2+4\eta/5+2\mu$&$3/2-\xi/2+4\eta/5+2\mu$\\
$\beta$ anti-structure bridge&$\xi/2-\eta/2$&$1/2+\eta/2$&$1/2+\eta/2$\\
$\beta$ vacancy bridge&&$1/2-3\xi/2+2\eta$&$1/2-3\xi/2+2\eta$
\end{tabular}
\caption{Activation energies (effective formation energy of initial state plus increase towards apex state) of the considered diffusion mechanisms in units of $E_0$. The expressions for the mirror cases ($\alpha\leftrightarrow\beta$) correspond to swapping the signs of $\eta$ and $\mu$.}\label{actenergies}
\end{table*}

\subsection{Estimating energetics}\label{diffusion_energetics}
Here the idea of a given diffusion mechanism's activation energy being the sum of non-interacting effective defect formation energies that was used qualitatively above will be explored quantitatively. This is the simplest conceivable model that takes into account the point defect energetics as described by the non-trivial parameters $\xi$ and $\eta$. Such a model has already been used by Kao and Chang\cite{kaointermet1993} for determining whether $\alpha$ or $\beta$ jump cycles dominate in a given system. In addition to equating the energy difference between apex and ground state with the activation energy as discussed in \ref{transstatetheory}, neglecting the interaction between neighbouring point defects and assuming a constant barrier height separating neighbouring stable states, it also considers only jumps of single atoms into nearest-neighbour vacancies as opposed to coupled jumps of several atoms. However, for the present purpose of determining the dominant mechanism this coupling will affect the energetics of all mechanisms and therefore cancel in the first approximation.

It is expedient to split the diffusion mechanisms' activation energy into the formation energy of the initial state and the increase in potential energy during the jump sequences. The former contribution follows from summing the respective effective defect formation energies from Eqs.~\eqref{effformenerg}, resulting in 
\begin{subequations}
\begin{align}
\eva+\evb&=E_0(1/2+\xi/2)\\
\eva&=E_0(1/4+\xi/4+2\eta/5+\mu)\\
2\eva&=E_0(1/2+\xi/2+4\eta/5+2\mu)\\
\eva+M\eab&=E_0(1/2+\eta/2)\\
\eab+M\evb&=E_0(-\xi+\eta).
\end{align}
\end{subequations}
for the divacancy, $\alpha$ jump cycles, $\alpha$ waltzing-step, $\beta$ anti-structure bridge, and $\beta$ vacancy bridge mechanisms, respectively. As the percolating mechanisms are to be considered only in the pertaining off-stoichiometric cases, the corresponding $\mu$ has already been chosen so that the effective formation energy of the $M$ necessary percolating defects vanishes, with $M$ a large number proportional to the system size. 

There are four elemental transitions the nearest-neighour jump mechanisms are made up of, viz.\ the transition of an A atom from its native lattice into a $\beta$ vacancy or back, and the analogous transitions for a B atom. The energy difference for the transition of an A atom from the $\alpha$ to the $\beta$ sublattice is given by $\eanb=\eva+\eab-\evb=E_0(1/2-\xi/2+\eta)$, similarly $\ebna=\evb+\eba-\eva=E_0(1/2-\xi/2-\eta)$. These energy differences will be used for the analysis of the diffusion mechanisms's intermediate states. Specifically, if $\eanb<0$ or $\eta<-1/2+\xi/2$ the generation of a $\beta$ antisite defect corresponds to an energy gain, analogously for an $\alpha$ antisite defect if $\eta>1/2-\xi/2$. 

For the divacancy mechanism, the energies of the intermediate states relative to the initial divacancy state are $\langle 0,\eanb,0,\ebna,0\rangle$. As a consequence, if $-1/2+\xi/2\le\eta\le1/2-\xi/2$ the divacancy state is the ground state, while a triple-defect state is the ground state if $\eta$ is outside of this interval. The apex state is always one of the two triple-defect state, with an energy difference to the initial state of $E_0(1/2-\xi/2+|\eta|)$.

The jump cycle will be considered in its form with an initial $\alpha$ vacancy for cycles with $K=4$ as is appropriate, e.g., for the B2 lattice. The sequence of energies are $\langle 0,\ebna,\ebna+\eanb,2\ebna+\eanb\rangle$, followed by a symmetrical descent. The second intermediate state is higher than the initial state by the positive energy needed to form an antisite pair, the same applies for the difference of the third to the first intermediate state. Therefore, only if $\eta<1/2-\xi/2$ the initial state with the single vacancy is the ground state and the third intermediate state the apex state so that this mechanism is properly called six-jump cycle. In the converse case, the first intermediate state is the ground state and the second intermediate state is the apex state, so that diffusion by this mechanisms corresponds to sequences of two-jump reorientations of the vacancy-antisite pair alternating with a four-jump cycle, which is more fittingly called 4+2-jump cycle, as discussed already above.

In the case of the waltzing-step mechanism with two initial $\alpha$ vacancies, the sequence of energies up to the state with mirror symmetry is $\langle 0,\ebna,\ebna+\eanb\rangle$. Either the first or the second intermediate state are the apex state depending on whether $\eta<-1/2+\xi/2$ or not. 

\begin{figure*}
\includegraphics{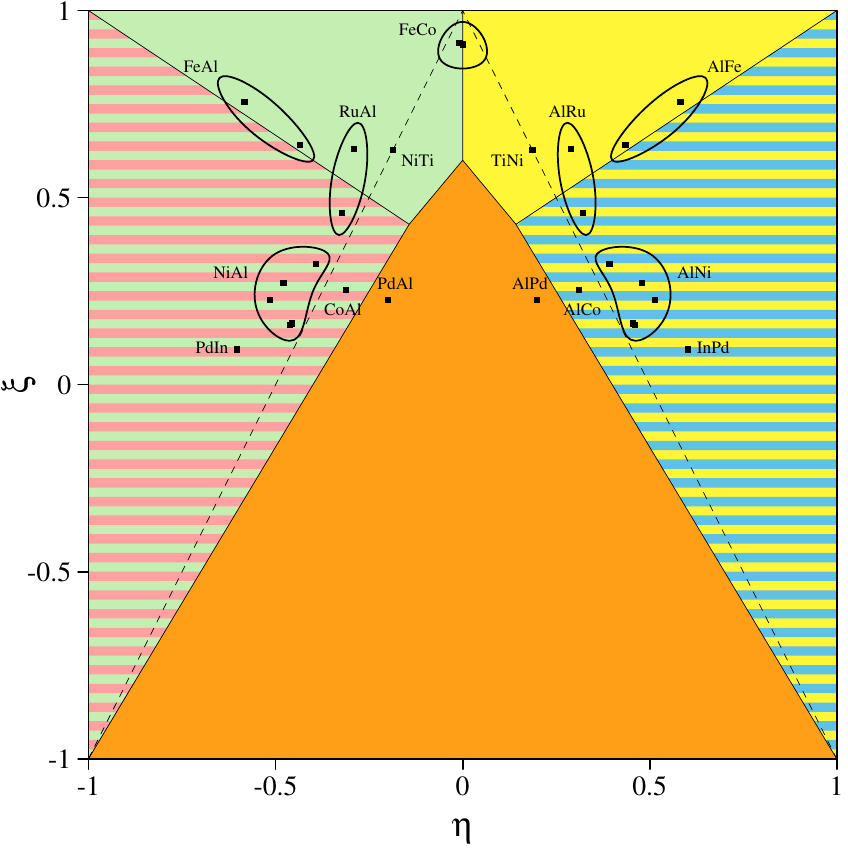}\qquad\includegraphics{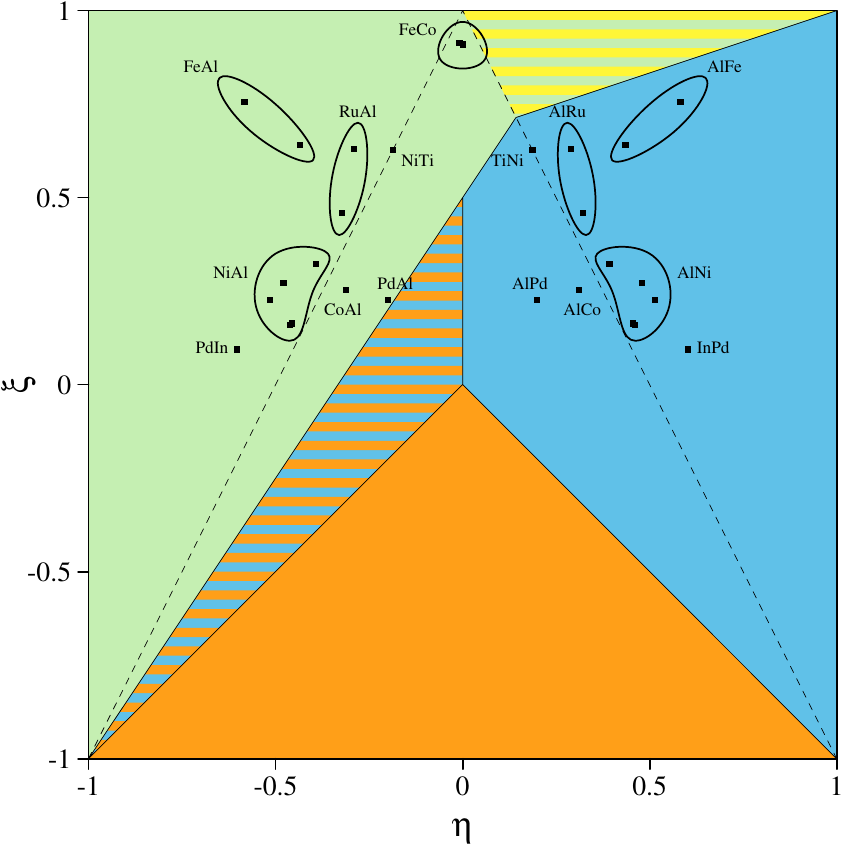}\\[.2cm]
\includegraphics{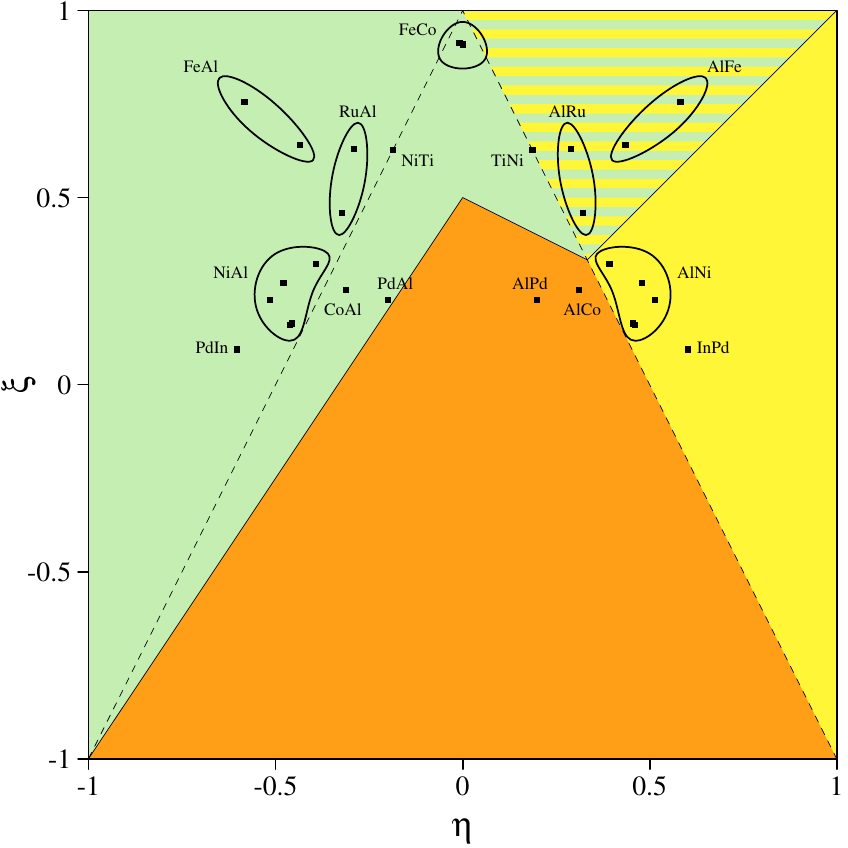}\qquad\includegraphics{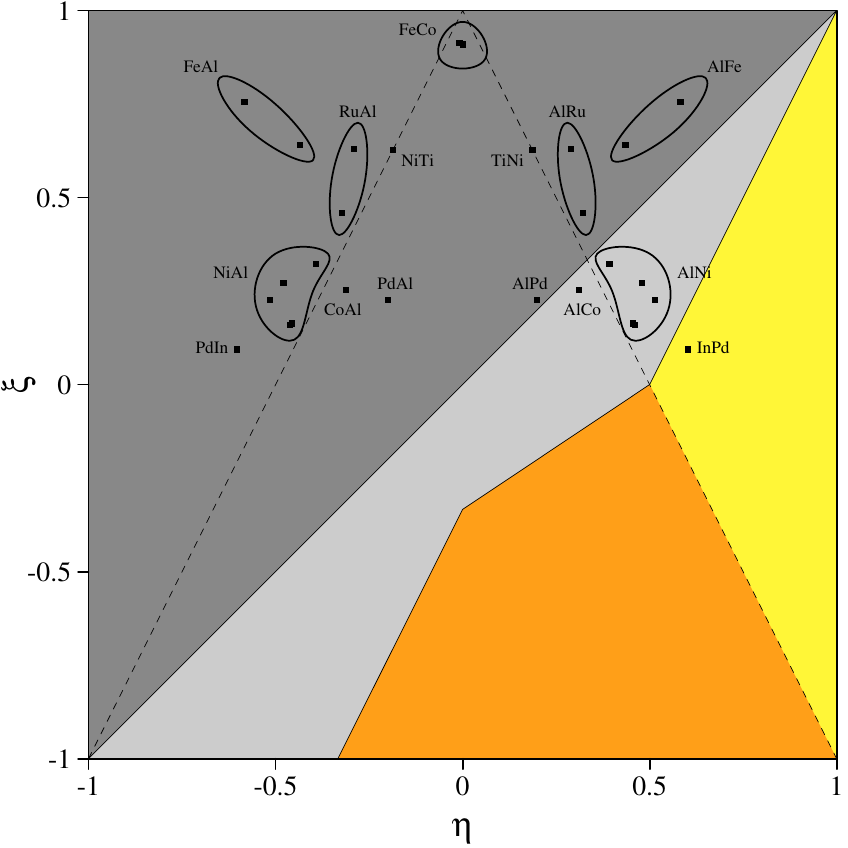}
\caption{Dominant diffusion mechanisms for non-interacting defect energetics. The divacancy mechanism is orange, jump cycles yellow ($\alpha$) and green ($\beta$), waltzing-step mechanisms pink ($\alpha$) and blue ($\beta$). Striped areas denote degeneracies. The dashed lines correspond to $|\eta|=1/2-\xi/2$, which decides on the apex states as discussed in the text. The left-top panel treats the stoichiometric case, the other three panels A-rich off-stoichiometry, namely B diffusion (right top), A diffusion without percolation of constitutional defects (left bottom), and A diffusion including percolating mechanisms (right bottom), where dark gray corresponds to the anti-structure bridge mechanism and light gray to the vacancy bridge mechanism.\label{phdg_mech}}
\end{figure*}

The anti-structure bridge mechanism under A-excess alternatingly passes through states with energy $0$ and $-\eanb$ relative to the initial state with an $\alpha$ vacancy, and either of those can be ground and apex states. For the vacancy bridge mechanism the energy difference of the intermediate state is $\eanb$, which is positive in the region of constitutional vacancies.

The apex state energies of the respective mechanisms and therefore their activation energies are tabulated in Tab.~\ref{actenergies}. The resulting dominant mechanisms defined by the smallest activation energies are illustrated in Fig.\ \ref{phdg_mech}. As in the case of point defects, also here all of the mechanisms considered above are active for some $(\xi,\eta)$, and furthermore for each mechanism there are representatives among the considered systems where it is expected to be active. The critical points are $(3/7,\pm 1/7)$ and $(3/5,0)$ for stoichiometry, $(0,0)$, $(1/2,0)$ and $(5/7,1/7)$ for B diffusion under A-excess, $(1/2,0)$ and $(1/3,1/3)$ for A diffusion, and $(-1,-1/3)$, $(-1/3,0)$ and $(0,1/2)$ for the vacancy bridge mechanism. Verifying the presented assignments of mechanisms to regions in $(\xi,\eta)$-space is conceptually simple but tedious and will not be performed here.

In the stoichiometric case, jump cycles and waltzing-step mechanisms share the minimal activation energy over a large region of the phase space, for instance $\beta$ jump cycles with the $\alpha$ waltzing-step mechanism in the A triple-defect region according to Fig.\ \ref{phdg_def} with $\eta<0$. This is because the apex states of the respective mechanisms agree up to a single point defect, which is an $\alpha$ vacancy in the case of the $\alpha$ waltzing-step mechanism and a $\beta$ antisite for $\beta$ jump cycles, having equal effective formation energies in the case of A triple defects. The waltzing-step mechanism, which leads only to diffusion of one component, is never exclusively active, so that diffusion of both constituents will be coupled. A point worth noting is that the major part of this region fulfills $|\eta|>1/2-\xi/2$, so that the corresponding jump cycles correspond to 4+2-jumps, with the ground state made up of a $\alpha$ vacancy and an $\beta$ antisite. On the other hand, the divacancy mechanism is active only for $|\eta|<1/2-\xi/2$, where it will take the form of the vacancy-leapfrog mechanism. Its dominance in the region of negative $\xi$ and small absolute values of $\eta$ can be easily understood, as small $\xi$ favour vacancies compared to the other mechanisms, which rely more on antisites, while $\eta$ shifts barrier height between the two triple-defect states, so that the smallest apex energy results from $\eta=0$.

In the off-stoichiometric case, the degeneracies between jump cycles and the waltzing-step mechanisms are lifted. Specifically for A-excess, for $\eta<0$ the $\beta$ jump cycles dominate, while for $\eta>0$ the $\beta$ waltzing-step mechanism dominates, causing diffusion of only the B minority component. This leads to the result that, with the exception of FeCo, all systems considered here show coupled diffusion for both constituents in the transition metal-rich case, while in the converse case the minority component (i.e., the transition metal) should diffuse with a smaller activation energy and therefore faster. 

It can be seen that for percolating constitutional defects, the anti-structure bridge mechanism is unconditionally responsible for majority diffusion, while the two antisite defects in the apex state of the vacancy bridge mechanism limit its effectiveness to a strip along the main diagonal. Still, according to the present model for all considered systems except for In-rich PdIn percolating mechanisms will dominate majority diffusion once they are made possible by sufficient deviations from stoichiometry.

There exists a region in the phase space where a percolation mechanism dominates majority diffusion, while the waltzing-step mechanism is responsible for minority diffusion. In this case, the information plotted in Fig.\ \ref{phdg_mech} cannot decide which component will have a smaller activation energy of diffusion. Pertinent calculations show that for the case of constitutional antisites, the anti-structure bridge mechanism corresponds always to the lower activation energy, while the vacancy bridge mechanism is lower in energy than the waltzing-step mechanism if and only if $\eta<1/4+\xi/4$.

\begin{figure}
\includegraphics{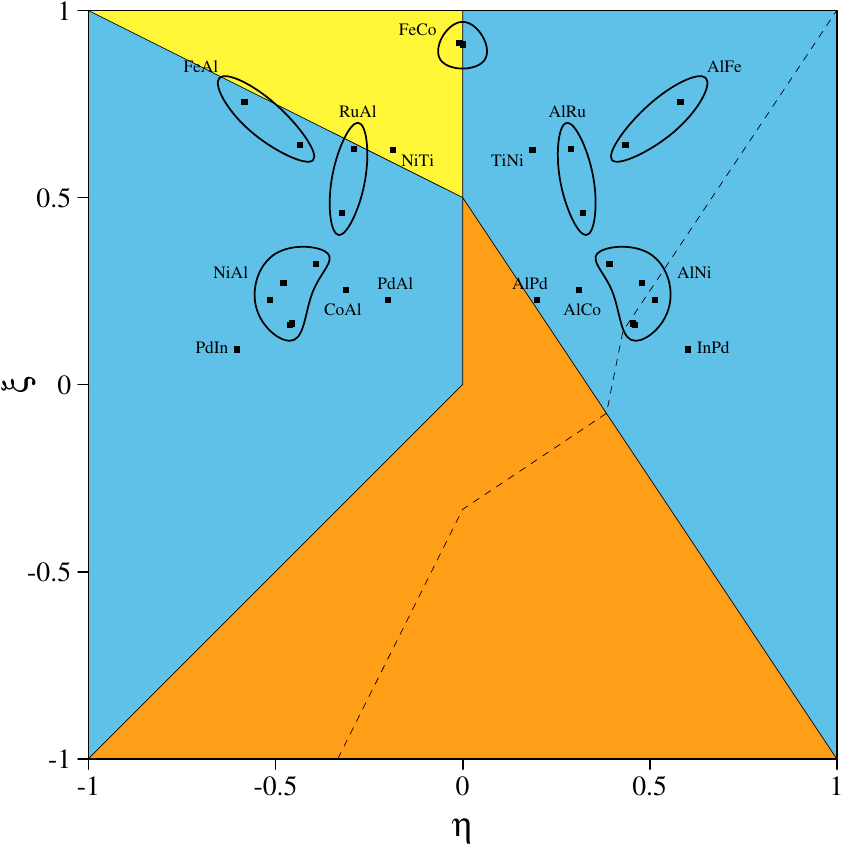}
\caption{Behaviour of activation energy of A diffusion as function of composition. Yellow denotes smaller activation energy for A-excess than for A-deficiency, blue the converse case, and orange equal activation energies for both cases. Active percolation mechanisms would enlarge the yellow region towards the dashed line.\label{vergl_EA_conc}}
\end{figure}

While above paragraphs discuss which component diffuses faster for given system and composition, the complementary question as to how the activation energy varies with composition is answered by Fig.\ \ref{vergl_EA_conc}. It shows that for major parts of the phase space the activation energy of diffusion for a given constituent is smaller if it is the minority species. Naturally, in systems where the divacancy mechanism is dominant, there is no effect of composition on the activation energy, as a divacancy is a neutral defect and its concentration is therefore independent of $\mu$. Only in a small region of the phase space a component's activation energy of diffusion drops with an increase in its concentration. However, once the specific percolation threshold is reached, the concerned component in most cases can diffuse with the smallest activation energy. The additional critical points are $(1/7,3/7)$ and $(-1/13,5/13)$.

\begin{figure*}
\includegraphics{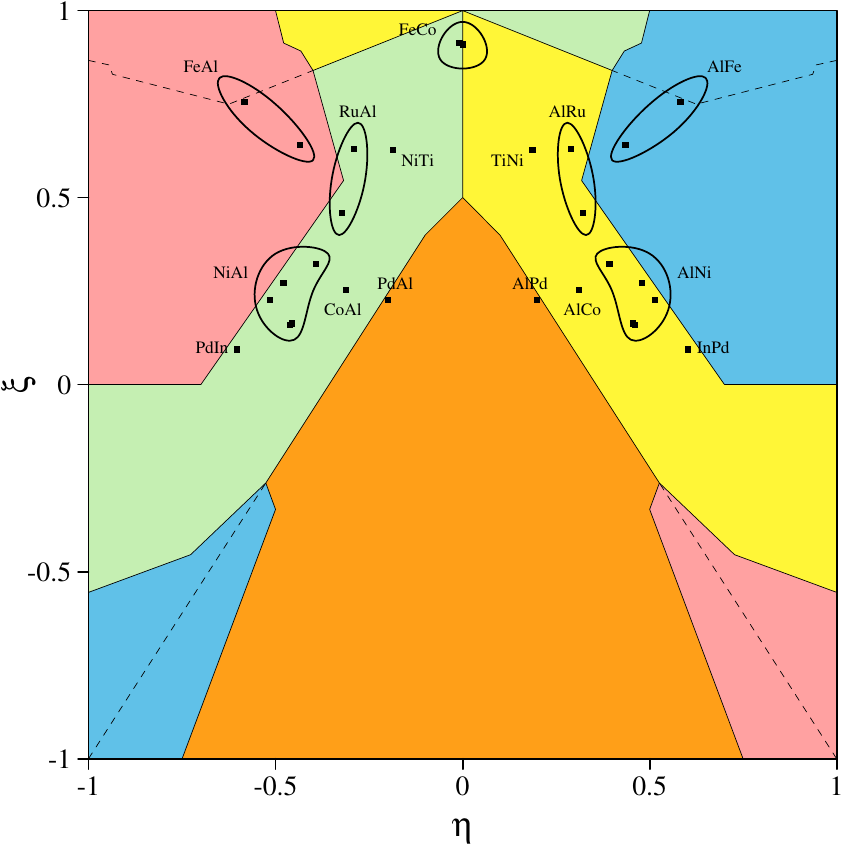}\qquad\includegraphics{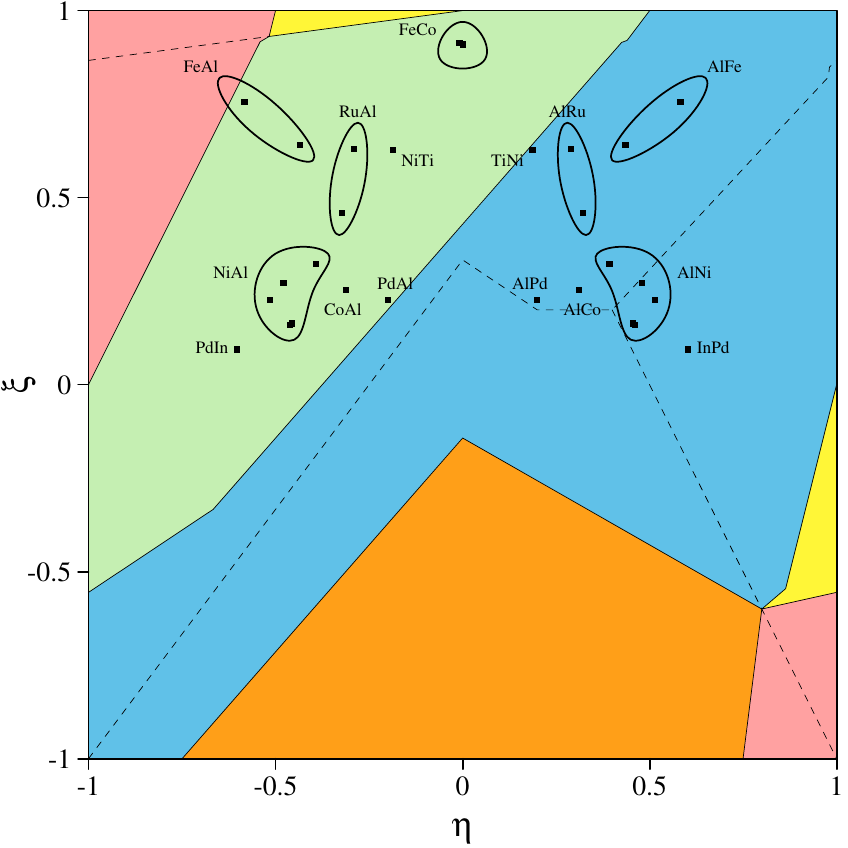}
\caption{Dominant diffusion mechanisms for pair-interaction energetics with $Z=8$ at stoichiometry (left) and A-excess (right). Colour code is the same as in Fig.\ \ref{phdg_mech}. Dashed lines delineate regions neglecting the waltzing-step mechanism, i.e., indicate the mechanism responsible for diffusion of the slower component.}\label{phdg_mech_interakt}
\end{figure*}

The proposed model of estimating activation energies for the different mechanisms is the simplest conceivable model that covers all mechanisms composed of atomic jumps into nearest-neighbour vacant sites. In reality, already in the small-temperature limit effects not treated here will complicate the picture. These include interactions between the point defects introduced during the various steps of the mechanisms as well as jump barriers separating the stable states that depend on the environment. As a consequence, for instance the three geometrically inequivalent six-jump cycles will have different saddle point energies (see the discussion in Sect.\ \ref{nial}). Further, in order to avoid especially costly intermediate states such as an aluminium vacancy in the case B2 transition metal-aluminides, the system can perform coupled jumps of several atoms. However, this would affect all mechanisms, so that for the present purpose of comparing the activation energies of different mechanisms it can be expected to cancel to a large amount. Finally, the interplay of different mechanisms due to equilibrium disorder at high temperature can in practice only be treated by Monte Carlo simulations (e.g.\ Ref.\ \onlinecite{xuprb2010}). In contrast to such simulations where a detailed understanding of the observed effects can be hard or even impossible to obtain, the reasoning presented here is rather transparent. In spite of its shortcomings as discussed above, the qualitative conclusions can be expected to be sound. Specifically, it motivates the existence of the jump mechanisms discussed and introduced in Sect.\ \ref{mechanisms}.

\subsection{Pair interactions}\label{diffusion_energetics_interakt}
In the model investigated in Sects. \ref{parametrization} and \ref{diffusion_energetics}, the energies of specific states depend only on the occupations of the distinct sublattices. In an interaction picture, it is the limit of a mean-field model for small defect concentrations and therefore corresponds to an infinite interaction range, distinguishing only between sublattices. In previous investigations (e.g., Ref.\ \onlinecite{athenesphilmaga1997}) often the opposite extreme of nearest-neighbour pair interactions has been considered, where the energies follow from summing the occupation-dependent contributions $e_{XY}$ of (undirected) nearest-neighbour pairs and the distinction of sublattices arises only due to spontaneous symmetry breaking of the thermodynamic ground state. Equating the formation energies of isolated point defects leads to the relations 
\begin{subequations}
\begin{align}
e_\text{AB}&=E_0(-1-\xi)/2Z\\
e_\text{AA}&=E_0(-\xi+\eta)/Z\\
e_\text{BB}&=E_0(-\xi-\eta)/Z,
\end{align}
\end{subequations}
where $Z$ is the coordination number. This implies that a pair-interaction description as conventionally used in Monte Carlo simulations has a priori the same generality as the model used here. However, allowing only a single vacancy in the simulation as is often done\cite{athenesphilmaga1997} corresponds to neglecting one degree of freedom and to potentially suppressing diffusion mechanism relying on the correlated movement of several vacancies, i.e., the divacancy and waltzing-step mechanisms.

Different from the model of non-interacting defect energetics, nearest-neighbour pair energetics lead to effective interactions between neighbouring point defects
\begin{subequations}
\begin{align}
\text{V$_\alpha$--V$_\beta$}:\quad&2e_\text{AB}=E_0(-1-\xi)/Z\\
\text{B$_\alpha$--A$_\beta$}:\quad&2e_\text{AB}-e_\text{AA}-e_\text{BB}=E_0(-1+\xi)/Z\\
\text{V$_\alpha$--A$_\beta$}:\quad&e_\text{AB}-e_\text{AA}=E_0(-1+\xi-2\eta)/2Z\\
\text{B$_\alpha$--V$_\beta$}:\quad&e_\text{AB}-e_\text{BB}=E_0(-1+\xi+2\eta)/2Z,
\end{align}
\end{subequations}
which are typically of attractive nature. These expressions specify how the energies of the diffusion mechanisms' intermediate states have to be modified for performing an analysis as done in Sect.\ \ref{diffusion_energetics}, but here for the assumption of nearest-neighbour pair energetics.

The resulting phase diagrams are given in Fig.\ \ref{phdg_mech_interakt}. As expected, the situation is more complicated than for non-interacting defect energetics, which is mostly a consequence of the cross-over from one apex state to the next not coinciding for the respective mechanisms due to the different impact of point defect interactions. The illustrated topology is representative for $Z>3$, i.e., for all relevant cases. Qualitatively, the introduction of local interactions (corresponding to $Z<\infty$) leads to a lifting of the degeneracies and the appearance of additional phase regions in the outer regions of the phase space. In the inner regions, the phase boundaries do not change much, however, so that the non-interacting classification (Fig.\ \ref{phdg_mech}) and the pair-interaction classification (Fig.\ \ref{phdg_mech_interakt}) agree on all considered B2 systems with the exception of Al diffusion in Al-rich PdAl and CoAl, which should proceed by the divacancy mechanism according to the non-interacting model, and by the six-jump cycle starting from the transition metal vacancy according to the pair-interaction model. Note that with local interactions neighbouring point defects can affect the energetics even if they do not participate in the jump, so that the percolating mechanisms cannot be assigned a definite activation energy. 


The fact that the dominant mechanisms as predicted by the opposite extremes of nearest-neighbour and infinite interaction range energetics agree so well gives support to the models. It shows that interactions between neighbouring point defect affect the energies of the intermediate states approximately equally for all mechanisms, so that these modifications cancel to a large extent. 




\section{Application to selected systems}
Here the predictions on the active diffusion mechanisms according to above models will be assessed by comparing to calculations and experimental diffusion data on NiAl and CoGa, two prototypical and well-investigated B2 systems.

\subsection{NiAl}\label{nial}
Both the non-interacting (Fig.\ \ref{phdg_mech}) and the pair-interaction model (Fig.\ \ref{phdg_mech_interakt}) predict Al 4+2-jump cycles to be the dominant mechanism of diffusion in all cases with the exception of Ni-deficiency, where the Ni six-jump cycles have comparable energy and the double Ni vacancy waltzing-step mechanism should allow faster Al-diffusion. Once the respective thresholds are reached, Ni-diffusion should proceed by the anti-structure bridge mechanism under large Ni-excess and Al-diffusion by the vacancy bridge mechanism under large Ni-deficiency. 

For NiAl there have been a number of calculations of diffusion energetics. For consistency, here the results reported by Xu and Van der Ven\cite{xufirstprinciples2009,xuintermet2009,xuprb2010} obtained by nudged elastic band density-functional theory calculations will be reported and used for deriving estimates for mechanisms that have not been considered yet. A consistent feature is the appearance of coupled jumps involving Al atoms. An Al vacancy is energetically expensive, therefore it is favourable that the vacated Al site is immediately reoccupied by another atom. If the effective translation of the vacancy is along $\langle 100\rangle$, the angle between the respective atomic translation vectors is acute in the idealized picture. Due to short-range repulsion between the jumping atoms, the Al site can therefore not be covered as effectively as for $\langle 110\rangle$ jumps with an obtuse angle or even collinear $\langle 111\rangle$ jumps. This is likely the reason for the observed increase of barrier energies for equivalent coupled jumps from $\langle 111\rangle$ over $\langle 110\rangle$ to $\langle 100\rangle$ as observed already in Ref.\ \onlinecite{mishinprb2003} and can be assumed to be of general validity.

For the Ni vacancy six-jump cycle Xu and Van der Ven give an apex energy of \quant{2.59}{eV} above the single vacancy state for the $\langle 110\rangle$ configuration, and higher values for the $\langle 100\rangle$ configurations or the equivalent mechanism with a second-nearest neighbour Al jump through the apex. Along with all other such calculations \cite{divinskiintermet2000,mishinprb2003,marinoprb2008} the Al vacancy six-jump cycle (or rather 4+2-jump cycle as will be seen) is not considered, which is explained either by its claimed notably higher activation energies \cite{divinskiintermet2000} or not justified at all. However, an estimate of its apex energy can be put together from those of its components: The nearest-neighbour pair of a Ni vacancy and a Ni anti-structure atom is \quant{0.6}{eV} lower in energy than an isolated Al vacancy, and the apex energy of this transition is \quant{0.35}{eV} with respect to the initial vacancy. Therefore the vacancy-antisite pair is the ground state, and it can perform its two-jump reorientations with \quant{0.95}{eV}. The four-jump part of the mechanism then proceeds by a coupled $\langle 110\rangle$ Al-Ni-jump into the Ni vacancy, which is just the first double jump of the $\langle 110\rangle$ Ni vacancy six-jump cycle with \quant{2.36}{eV}. However, the presence of the Ni anti-structure atom can be expected to decrease this energy further, just as it decreases the barrier energy of the coupled Al-Al-jump from \quant{1.01}{eV} in the isolated case to \quant{0.53}{eV} in the Ni six-jump cycle. Therefore, the activation energy of the Al 4+2-jump cycle can be expected to be \quant{1.76}{eV} above the isolated Al vacancy, probably decreased by a few tenths of eV. Note that while the four-jump part as described here leads to a $\langle 110\rangle$ displacement of the vacancy, in the six-jump description it would correspond to a $\langle 100\rangle$ cycle.

The divacancy mechanism should take the triple-defect modification with the Ni triple defect as ground state. For the $\langle 100\rangle$ jump its apex energy during the Al jump is \quant{0.75}{eV} above the nearest-neighbour divacancy, the energy of which should not deviate by more than \quant{0.1}{eV} from isolated vacancies\cite{korzhavyiprb2000}, while the energy is much higher for other jump geometries. For the double Ni vacancy waltzing-step mechanism the energetics have not been computed yet, too. Again the coupled $\langle 110\rangle$ Al-Ni-jump with \quant{2.36}{eV} can be considered as an estimate for its energy with respect to the double Ni vacancy, with perhaps slightly smaller energy for $\langle 111\rangle$ jumps but \quant{2.88}{eV} for $\langle 100\rangle$ jumps. The presence of the second Ni vacancy will probably affect these barrier energies to a small degree. The converse case of an Al waltzing step corresponds to the \quant{0.75}{eV} of an Al jump into a divacancy relative to the effective formation energy of the complex of a divacancy and a Ni anti-structure atom. To compare, the direct second-nearest neighbour jumps within one sublattice cost \quant{2.76}{eV} for Ni and \quant{1.49}{eV} for Al, which shows that the availability of vacancies with low or vanishing effective formation energies decreases the energy for direct jumps within a sublattice by a non-jumping atom withdrawing into the vacancy, which is the essence of the waltzing-step mechanism. 

The barrier energy for the Ni anti-structure bridge mechanism is around \quant{1}{eV} relative to the Ni vacancy, depending on the surrounding configuration of the constitutional antisites. The Al vacancy bridge mechanism proceeds via coupled jumps with \quant{1.65}{eV} for $\langle 100\rangle$ antisite translations and \quant{1.01}{eV} for $\langle 110\rangle$, again probably even less for $\langle 111\rangle$. Constitutional defects show repulsions of about \quant{0.1}{eV} over $\langle 100\rangle$, while interactions over farther distances are negligible\cite{korzhavyiprb2000}. With the resulting suppression of $\langle 100\rangle$ coordinations the probabilities for $\langle 110\rangle$ and $\langle 111\rangle$ pairs will rise, so that the respective percolation thresholds of $c_\text{Ni}=0.55$ and $0.475$ will probably be even smaller.

As Xu and Van der Ven do not report their defect formation energies, an average of the values in Tab.\ \ref{compilation} with \quant{E_0=4.14}{eV}, $\xi=0.228$ and $\eta=-0.460$ will be assumed in the following. Computing the activation energies of the various mechanisms by adding the corresponding effective formation energies for the necessary vacancies to the barrier energies gives the following hierarchy: For Ni-deficiency, the Al vacancy bridge mechanism has an activation energy of \quant{1.97}{eV}, followed by the Ni waltzing-step mechanism with about \quant{2.36}{eV}, the Ni six-jump cycle with \quant{2.59}{eV} and direct Ni jumps to second-nearest neighbours with \quant{2.76}{eV}. Further options like the triple-defect mechanism have \quant{3.29}{eV} or higher and can therefore safely be neglected. For Ni-excess, the Ni anti-structure bridge mechanism is lowest in energy with \quant{2.11}{eV}, followed by direct Al second-nearest neighbour jumps with \quant{2.91}{eV} and the Al 4+2-jump cycles with a few tenths of eV below \quant{3.18}{eV}. The triple-defect and Al waltzing-step mechanisms with \quant{3.29}{eV} are already clearly higher. For stoichiometry, the energies of Ni six-jump cycles, Al 4+2-jump cycles, the triple-defect mechanism, and direct Al jumps all cluster closely around \quant{3.3}{eV}. These values reproduce the rankings of the simple models of Sects.\ \ref{diffusion_energetics} and \ref{diffusion_energetics_interakt}, only the dominance of the Ni six-jump cycle for Al diffusion at Ni-deficiency due to the inexpensive coupled Al-Al-jumps is missed, and the second-nearest neighbour jumps are obviously not included.

The experimental evidence in this system convincingly demonstrates constant Ni tracer diffusivities as a function of composition for Ni-deficiency and smoothly rising diffusivities with Ni-excess. This is mirrored in the activation energies of Ni diffusion with a constant value of \quant{3.0}{eV} on the Ni-deficient side and sharply dropping over about 53 \% Ni content \cite{frankactamat2001}. Assuming In diffusion to mirror Al tracer diffusion\cite{minaminodefdiffforum2001}, it shows a sharp minimum in diffusivity and a maximum in activation energy around stoichiometry. In contrast, the simulated diffusivities for both components \cite{xuprb2010} show a sharp minimum at stoichiometry, a steep increase towards Ni-excess, and a smaller increase with equal diffusivities of Ni and Al towards Ni-deficiency. 

A number of open questions remain for diffusivity in this system: First, the concentration of the bound divacancy, being a composition-conserving excitation, is nearly independent of composition\cite{xuintermet2009}. This will also hold for the migration energy of the Al jump into the divacancy, therefore the contribution of the triple-defect to diffusion is independent of composition. The decomposition of diffusivities in Ref.\ \onlinecite{xuprb2010} shows that when going from stoichiometry to Ni-deficiency, as expected, the Al vacancy bridge jumps followed by Ni six-jump cycles and the Ni waltzing-step (which remained unrecognized and probably classified as coupled Ni-Al back and forth jumps) overtake the triple-defect mechanism and therefore yield a higher contribution to diffusion in absolute values. As the composition is in the region where the vacancy bridge mechanism should be able to contribute to long-range diffusion, it is surprising that specifically the simulated Al diffusivity does not grow much compared to the stoichiometric case, and also the Ni diffusivity is comparable to the values at small Ni-excess. For larger Ni-excess the Al 4+2-jump cycles should become active, with its two-jump reorientation classified as Ni back and forth hops and the four-jump cycle again as coupled Ni-Al-hops. However, the expected contribution from direct Al second-nearest neighbour jumps remain unaccounted for.

On the other hand, the experimental finding of smoothly rising Ni and Al diffusivities with Ni-excess can be explained by the Ni anti-structure bridge jumps, direct Al jumps, and Al 4+2-jump cycles, all of which will rise in frequency with Ni-excess due to entropic reasons. Specifically the latter mechanism has hitherto been overlooked\cite{frankactamat2001,herzigintermet2004}. While in the Ni-deficient case the rising Al diffusivity will probably be due to the Al vacancy bridge mechanism, either solely responsible or enhancing the Al diffusivity in Ni six-jump cycles, the apparently constant Ni diffusivity and activation energy is perhaps only an accidental effect and can arise from a gradual shifting from an anti-structure bridge assisted Al 4+2-jump cycle to a combination of Ni waltzing-step mechanism, Ni six-jump cycles and direct Ni second-nearest neighbour jumps, as due to high-temperature disorder the chemical potential $\mu$ and therefore the point defect concentrations show a smooth behaviour \cite{xuintermet2009}. This seems plausible as all of these three mechanisms are expected to have a lower activation energy than the triple-defect mechanism assumed to be responsible previously\cite{frankactamat2001,herzigintermet2004}, which will more than balance their entropic disadvantage due to the reliance on constitutional vacancies. For the relation between the high-temperature experimental activation energies and zero-temperature calculations see also the following section.

\subsection{CoGa}
This compound is the prototypical example for a system where diffusion is assumed to proceed via the triple-defect mechanism. In tracer diffusion measurements over a large temperature range, deviations from linearity in the Arrhenius plots of both constituents' diffusivities were observed\cite{stolwijkphilmaga1980}. It turned out that the experimental data could be well fitted assuming two or more activation energies per component and composition, ranging from \quant{2}{eV} to \quant{5}{eV}, with the additional restriction that the larger Co activation energy be equal to the smaller Ga activation energy. This was taken as evidence for the simultaneous operation of several distinct diffusion mechanisms, where the one corresponding to the coupled activation energies was postulated to be the triple-defect mechanism\cite{stolwijkphilmaga1980} in analogy to the case of pure elements, where a contribution from divacancies was assumed to be responsible for deviations from linear Arrhenius plots.

However, also the contrasting opinion that anharmonicity of a single kind of defect can be responsible for the observed curvatures has long been held\cite{gilderprb1975}. Only recently the analogous problem of curvatures in the Arrhenius plots of vacancy concentrations has been settled by ab-initio calculations taking anharmonicity fully into account\cite{glenskprx2014}, where it has been shown that the formation entropy in cases as simple as single Al and Cu vacancies depends significantly on temperature, being able to fully explain the experimentally observed curvatures.

In the case of an ordered compound, a temperature-dependent order parameter can be a further source of curvature.\cite{athenesphilmaga1997} In the light of this, the initial justification for the triple-defect mechanism to be dominant at specific temperature ranges in CoGa seems doubtable, given also the fact that the respective diffusivity prefactors $D_0$ of the multi-exponential fits span ten orders of magnitude. Indeed, fitting the diffusion data\cite{stolwijkphilmaga1980} by a single mechanism that has an entropy varying linearly with temperature leads to values of $\chi^2$ per degree of freedom that are practically equal to those of the multi-exponential fits. For the Co-rich and stoichiometric compositions, both constituents' diffusion energies range from \quant{2.0}{eV} to \quant{2.7}{eV}, while the fitted values become physically implausible at Al-rich compositions. This shows that at least for stoichiometric and Co-rich compositions the observed curvature cannot be taken as indication of a crossover between two diffusion mechanisms. 

In default of point defect energy calculations it seems safe to assume that the situation here corresponds most closely to NiAl, also displaying constitutional transition-metal vacancies albeit with a smaller $E_0$ leading to higher disorder at comparable temperatures\cite{neumannactamet1980} and a wide region of B2 phase stability. Furthermore, also the experimental behaviour of the diffusivities conforms, with Ga showing faster diffusion on either side of stoichiometry, while the Co diffusivity seems unaffected by Ga-excess. The diffusion acceleration due to deviations from stoichiometry is smaller than in the case of NiAl, probably because of the higher degree of disorder. As in NiAl, contrary to expectations there is no discernible setting-in of the anti-structure bridge mechanism at some Co-rich percolation threshold, rather this mechanisms seems only to play a secondary role in enhancing the Co diffusivity effected by some other mechanism, active on both Co and Ga, with progressive Co-excess. The reason for this probably again lies in the comparatively low degree of order, which agrees with Co diffusion as measured by quasi-elastic neutron scattering being well described by uncorrelated jumps between both sublattices\cite{kaisermayrprb2001}. On the other hand, the fact that Ga diffuses decidedly faster than Co in Co$_{45.2}$Ga$_{54.8}$ at low temperatures is an even clearer indication for the vacancy bridge mechanism than in NiAl, which is probably the most relevant feature displayed by the available diffusion data in this compound.

\section{Conclusions}
To conclude, in this article a complete characterization of the defect thermodynamics in binary systems with two chemically inequivalent sites has been given. The various possible combinations of constitutional defects and thermal excitations have been enumerated, and specifically the relation of constitutional defects and thermal triple-defect excitations has been clarified. Published computations of point defect energetics have been classified accordingly, and it has been demonstrated that all the theoretical possibilities have representatives among actual binary compounds. 

Further, a list of jump mechanisms composed of nearest-neighbour translations of atoms into vacancies has been given, and its exhaustiveness for small defect concentrations has been motivated. These mechanisms are the well-known six-jump cycles and divacancy mechanism, while in addition the waltzing-step mechanism has been proposed here for the first time. For percolating clusters of antisites and vacancies, the anti-structure bridge and the as yet little known vacancy bridge mechanisms, respectively, have been quoted, along with precise values for the appropriate percolation thresholds. 

Two models for computing diffusion energetics directly from point defect energies have been considered, one in terms of non-interacting defects and the other for nearest-neighbour pair interactions. As for the point defect thermodynamics, a classification according to dominating diffusion mechanisms at low temperatures has been given for both models. The close agreement between their predictions is judged as a strong argument for their qualitative correctness. Specific findings are that the hitherto not considered waltzing-step mechanism will dominate minority diffusion over a large range of systems, and that the six-jump cycles will in most cases actually proceed as 4+2-jump cycles. This has the consequence that for transition metal-group III compounds such as NiAl the six-jump cycles starting from the main group vacancies are expected to dominate, contrary to prevailing opinion. Also, the divacancy mechanism is expected to be less relevant for strongly asymmetric systems, where it would take the triple-defect form, than for symmetric systems.

Finally, the model predictions have been compared to computed diffusion energetics in NiAl, and satisfactory agreement has been demonstrated. These findings lead to new interpretations of experimentally determined composition-dependent diffusivities in NiAl and CoGa.

\section*{Acknowledgements}
My interest in point defects and diffusion in compounds has been inspired by the history of research in my previous group at the University of Vienna, and the arguments presented here have been formed during continual scientific exchange especially with Bogdan Sepiol, Markus Stana, and Gero Vogl. This work was supported by the Austrian Science Fund (FWF): P22402 and by the Deutsche Forschungsgemeinschaft (DFG) through TRR 80.

\bibliography{\bibspath abkuerz,\bibspath dummy,\bibspath curved_arrh,\bibspath diff,\bibspath punktdefekttheorie,\bibspath defektenergien,\bibspath nial,\bibspath unsereneutronen}
\end{document}